\documentclass[reprint,bibnotes,amsmath,amssymb,aps,prb,showpacs,floatfix,superscriptaddress,longbibliography]{revtex4-1}

\usepackage{amsmath}
\usepackage{amsfonts}
\usepackage{amssymb}
\usepackage{amsxtra}
\usepackage{xcolor}
\usepackage{graphicx}
\usepackage{subfigure}
\usepackage{dcolumn}
\usepackage{float}
\usepackage{bm}
\usepackage[breaklinks=true,colorlinks,citecolor=blue,linkcolor=blue,urlcolor=blue]{hyperref}

\newcommand{\nn}{\nonumber}

\newcommand{\bsigma}{\boldsymbol{\sigma}}
\DeclareMathAlphabet{\bi}{OML}{cmm}{b}{it}
\def\be{\begin{equation}}
\def\ee{\end{equation}}
\def\bearr{\begin{eqnarray}}
\def\eearr{\end{eqnarray}}
\def\la{\langle}
\def\ra{\rangle}

\begin{document}
\title{Nonlinear optical conductivity of a generic two band systems, with application to doped and gapped graphene}
\author{Ashutosh Singh}
\affiliation{Department of Physics, Indian Institute of Technology Kanpur, Kanpur - 208016, India}
\author{Tuhina Satpati}
\affiliation{Department of Physics, Indian Institute of Technology Kanpur, Kanpur - 208016, India}
\author{Kirill I. Bolotin}
\affiliation{Department of Physics and Astronomy, Vanderbilt University, Nashville, TN, USA}
\author{Saikat Ghosh}
\affiliation{Department of Physics, Indian Institute of Technology Kanpur, Kanpur - 208016, India}
\author{Amit  Agarwal}
\email{amitag@iitk.ac.in}
\affiliation{Department of Physics, Indian Institute of Technology Kanpur, Kanpur - 208016, India}

\date{\today}

\begin{abstract}
We present a general formulation to calculate the dynamic optical conductivity, beyond the linear response regime, of any electronic system  whose quasiparticle dispersion is described by a two band model. Our phenomenological model is based on the optical Bloch equations. In the steady state regime it yields an analytic solution for the population inversion and the interband coherence,  which are nonlinear in the optical field intensity, including finite doping and temperature effects. We explicitly show that the optical nonlinearities are controlled by a single dimensionless parameter 
which is directly proportional to the incident field strength and inversely proportional to the optical frequency. 
This identification leads to a unified way to study the dynamical conductivity and the differential transmission spectrum across a wide range of optical frequencies, and optical field strength. 
We use our formalism to analytically calculate the nonlinear optical conductivity of doped and gapped graphene, deriving the well known universal ac conductivity of $\sigma_0={e^2}/4\hbar$ in the linear response regime of low optical intensities (or equivalently high frequencies) and non-linear deviations from it which appear at high laser intensities (or low frequencies) including the impact of finite doping and band-gap opening.
\end{abstract}

%
\maketitle
\section{Introduction}
Since it's discovery, graphene, a truly two-dimensional system, has been at the forefront of material research [\onlinecite{Novoselov22102004,Nature_438_197(2005), 10.1038/nmat1849}]. On account of its linear dispersion and high carrier mobility, graphene has demonstrated remarkable electronic and optical properties [\onlinecite{katsnelson2012graphene}]. In the linear response regime, with weak optical field induced momentum linearly coupling  to the charge carriers in graphene, spectacular physical effects have been predicted and observed which are significantly enhanced or peculiar, when compared to their bulk counterparts. An early such surprise was the observation of a strong coupling of a single monolayer of carbon atoms to electro-magnetic radiation with almost a constant absorption coefficient of $2.3\%$ over a broad range of optical frequencies.  The corresponding optical conductivity is elegantly expressed in terms of  universal constants, in the form $\sigma (\omega)= \sigma_0 \equiv {e^2}/{4\hbar}$ [\onlinecite{apl131905,Nair2008,PRL.100.117401,nphys989, Mak24082010}]. Several exotic predictions and observations followed, including ultra-high mobility in pristine graphene [\onlinecite{Bolotin2008}], Klein tunneling 
[\onlinecite{PRL_102_026807(2009)}], weak localization [\onlinecite{PRL_100_056802(2008)}] and quantum hall effect 
[\onlinecite{Nature_438_197(2005), Science_315_1379(2007), Nature_462_196(2009)}]. 

Following the first prediction of  universal optical conductivity for graphite honeycomb lattices [\onlinecite{Ando2002}], several theoretical works  extended the formulation to graphene using the Dirac cone approximation for clean samples [\onlinecite{PhysRevB.73.245411,Gusynin2006,PRB.75.165407}], and with disorder [\onlinecite{PRB.73.125411}]. Experimental observation of the universal optical conductivity [\onlinecite{apl131905,Nair2008,PRL.100.117401,nphys989, Mak24082010}] gave a strong impetus to the field  and motivated further theoretical work. These include studying the bandstructure effects on optical conductivity beyond the Dirac cone approximation [\onlinecite{Stauber_TB}], analyzing effect of strain [\onlinecite{EPL.92.67001}], role of substrate [\onlinecite{EPL.84.38002}] and the effect of electron-electron interactions on the optical conductivity [\onlinecite{PRL.98.216801, PRL.100.046403, PRB.78.085416, EPL.83.17005}].  In addition to the linear conductivity, several remarkable non-linear optical effects in graphene have also been explored [\onlinecite{Zhang1, arXiv1106.4838, Mishchenko, arXiv150600534M, 0295-5075-79-2-27002, 0295-5075-84-3-37001, EPL.84.38002, PhysRevB.82.201402, JPCC_Mikhailov, Ruvinskii, arXiv150901209}], motivating a plethora of applications[\onlinecite{nphoton.2010.186, Nano.nn300989g, IEEE_Avouris, nphoton.2012.262, nl102824h}] based on broadband nonlinear optical properties of graphene.

In particular, non-linear optical response along with linear dispersion in graphene imply higher harmonic generations, and the large velocities of carriers results in highly efficient electron-photon coupling. Furthermore, gapless excitation leads to effective resonant non-linear excitation in graphene. These interesting possibilities have motivated a plethora of studies based on non-linear response of these 2-d materials, ranging from microwave [\onlinecite{Gusynin2006}], terahertz [\onlinecite{Mikhailov2009}] to optical frequencies [\onlinecite{nphoton.2010.186}]. Such response has led to second and third harmonics generation in the optical [\onlinecite{Dean2009,Glazov2011}] and THz domain [\onlinecite{Crosse}], frequency mixing ranging from microwave to optical excitations [\onlinecite{PRL_105_097401(2010), Nat_Phot_6_554(2012)}], self-phase modulation and optical Kerr effects [\onlinecite{Nano_Lett_11_5159(2011), Chu2012104}], photon drag [\onlinecite{PhysRevB.81.165441}], THz driven chiral edge photo-currents [\onlinecite{PhysRevLett.107.276601}] or, dynamic Hall-effect driven by circularly polarized optical frequencies [\onlinecite{PRL_105_227402(2010)}]. Such unusual effects triggered a range of applications including graphene based rf modulators [\onlinecite{Nature_474_64(2011)}], optically gated transistors [\onlinecite{Nat_Nano_7_363(2012), Nano_Lett_14_1242(2014)}], photo-detectors [\onlinecite{Nat_Nano_9_780793(2014)}], graphene saturable absorbers for mode-locking [\onlinecite{APL_96_111112(2010)}] and even proposal for nonlinear interaction at the level of single photons [\onlinecite{NJP_17_083031(2015)}].

\begin{figure}[t!]
\centering
\includegraphics[width=.8\linewidth]{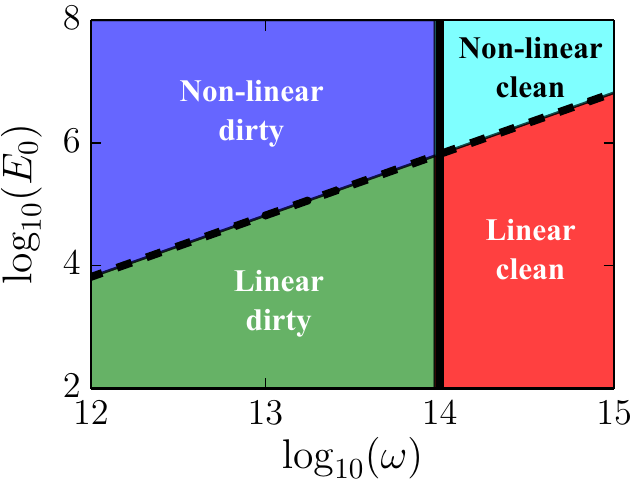}
\caption{ Classification of different optical response regimes in the parameter space of optical frequency (in Hz) and the incident radiation field strenth (in units of V/m). The  vertical black line marks the $\omega=\gamma_2$ boundary between the dirty or low frequency (left) and the clean or high frequency (right) limits. The black dashed line marks the  $\zeta \equiv e v_{\rm F} E_0/(\hbar \omega \sqrt{\gamma_1 \gamma_2)} =1$ line which is the boundary of the linear (bottom) and non-linear (top) response regimes.  Note that for a given frequency and material (with fixed damping parameters) the linear to non-linear regime crossover occurs at smaller field strength.  Here we have chosen $v_{\rm F} =10^6$ m/s, and $\{\gamma_1,\gamma_2\} = \{10^{12},10^{14}\}$ Hz, based on parameters for graphene (see Ref.~[\onlinecite{Zhang1}] and the references therein).
\label{fig1}}
\end{figure}

At a fundamental level, non-linear response functions serve as excellent tools for probing intrinsic material properties. This provides a rich class of information on material symmetry and selection rules, intricacies of band-structure, electron-spin relaxation and decoherence mechanisms that are otherwise hidden in the linear response regime. There is thus significant motivation to develop a unified theoretical framework to address non-linear response of low dimensional systems in general, that is applicable over a range of optical excitation frequencies, 
and a wide range of the incident optical field strength. 

Motivated by Mishchenko~[\onlinecite{Mishchenko}], in this article we present a theoretical framework for calculating non-linear optical conductivity of a general two-band system, which is   applicable over a large range of optical frequencies and field strengths. We recast the wave-function based approach of Ref.~[\onlinecite{Mishchenko}], into a density matrix based approach and this allows us to incorporate the impact of finite temperature, finite doping (chemical potential) etc. into the non-linear conductivity calculations.
In particular, we provide steady state solutions for the coupled Maxwell-Bloch equations for a two band system, in which we include electron-electron and electron-phonon scattering phenomenologically via the inter-band population inversion decay rate $\gamma_1$ and the coherence decay rate $\gamma_2$. The steady state population inversion and the inter-band coherence 
is then used to calculate the optical conductivity for a generic two band systems, and the optical transmission spectrum for a two dimensional system in general. We apply the developed formulation to analyze the non-linear optical conductivity of doped and gapped graphene in detail. However, the formalism can also be applied to other systems such as bilayer graphene, Weyl semimetals, phosphorene etc. whose low energy electronic properties are captured by a 
two band model. 

A natural outcome of our formalism is that, with $\gamma_1$ and $\gamma_2$  as two phenomenological input parameters, it allows us to classify the parameter space in terms of the incident optical frequency ($\omega$) and field strength ($E_0$) in four regimes: a) linear response in the clean regime, b) linear response in the dirty regime, c) non-linear response in the clean regime and finally d) the non-linear response in the dirty regime [see Fig.~\ref{fig1}]. 
What we call the clean (dirty) regime can also be called the collision-less or high frequency limit (collisional or low frequency), and is quantified by the region $\omega \ge \gamma_2$ ($\omega \ll \gamma_2$). 

The linear ($\zeta <1$) or nonlinear ($\zeta >1$) response of the system is quantified by a single dimensionless parameter [\onlinecite{Mishchenko}],
\be \label{zeta}
\zeta \equiv \frac{ e v_{\rm F} E_0}{\hbar \omega \sqrt{\gamma_1 \gamma_2}}~, 
\ee where $v_{\rm F}$ denotes a material dependent effective velocity.  Non-linear optical effects start becoming dominant  either on increasing the field strength keeping the frequency constant, or alternatively by decreasing the frequency while keeping the field strength constant. 

The manuscript is organized as follows: in Sec.~\ref{sec2}, we describe a general two band systems and its optical response via the population inversion and coherence. This is followed by a discussion of the non-linear optical conductivity in Sec.~\ref{sec3}, and the differential transmission of a freestanding two dimensional material in Sec.~\ref{sec4}. In Sec.~\ref{sec5} and Sec.~\ref{sec6}, we discuss the dynamic conductivity of doped and gapped graphene respectively, in various limiting cases of linear response in clean limit (previously known), linear response in dirty limit, non-linear response in the clean limit and the most general case of non-linear response in the  dirty limit. Finally we summarize our findings in Sec.~\ref{sec7}. 

\section{Population inversion and coherence in a general two band system}
\label{sec2}

We start with a very generic two band electronic system, in presence of an electromagnetic radiation.
The electromagnetic field is treated classically in the coulomb gauge, with the vector potential ${\bf A}$ satisfying  ${\bf \nabla}\cdot {\bf A} = 0$ and the scaler potential $\Phi = 0$,  yielding $\bf B = {\bf \nabla} \times {\bf A}$ and ${\bf E} = - \partial_t {\bf A}$.
The Hamiltonian describing the dynamics of an electron in presence of an external electromagnetic field, is given by $H_{ \rm em} = H_0 (\hbar \hat{\bf k} \to \hbar \hat{\bf k} 
+ e{\bf A} )$, which is usually approximated as  $H_{\rm em} \approx H_0 + e 
\hbar^{-1}{\bf A} \cdot\nabla_{\bf k} H_0 $. Note that while $H_{\rm em} $ is an approximation for general two band systems, it is exact for systems described by the two dimensional Dirac Hamiltonian, for which $H_0$ only depends linearly on the wave-vectors. Furthermore, from the perspective of calculating optical conductivity, this is akin to neglecting the diamagnetic part of the current (which anyway vanishes for Dirac systems since $\partial^2 H_0/\partial k_i^2 = 0$), and only focussing on the paramagnetic part of the response function [\onlinecite{Stauber_NJP}].

In the eigen-basis of $H_0$, the effective Hamiltonian, 
can be rewritten as, $H_{\rm em} = H_0' + \bf A \cdot {\bf M}$, 
where $H_0'$ is a diagonal matrix comprising of the dispersion of two bands. The elements of the matrix 
${\bf M}$ are defined by  
${\bf M}^{\lambda \lambda'} ({\bf k}) = e \hbar^{-1} \la \psi^\lambda |{\nabla_{\bf k}} H_0 |  \psi^{\lambda'} \ra$. 
We consider light to be incident perpendicular to the sample, with negligible transverse momentum, such that  it does not significantly alter the electron momentum. This sets a selection rule, 
allowing only vertical transitions in the momentum space.
More explicitly, $H_{\rm em} = H_0' + H_I$,
where we have 
\be
H_0' =  \sum_{\bf k} \varepsilon_{\bf k}^c {a^c_{\bf k}}^\dagger a^c_{\bf k} + \varepsilon_{\bf k}^v {a^v_{\bf k}}
^\dagger a^v_{\bf k}~,
\ee
with  ${a^c_{\bf k}}({a^{c\dagger}_{\bf k}}) $ and ${a^v_{\bf k}}({a^{v\dagger}_{\bf k}})$ being the annihilation 
(creation) operator for electron in the conduction and valance band respectively. The interaction part of the Hamiltonian is given by 
\be 
\frac{H_I}{\hbar} =  \sum_{\bf k}  \Omega^{cc}_{\bf k} {a^c_{\bf k}}^\dagger a^c_{\bf k} + \Omega^{vv}_{\bf k} 
{a^v_{\bf k}}^\dagger a^v_{\bf k} + \Omega^{cv}_{\bf k} {a^c_{\bf k}}^\dagger a^v_{\bf k} + \Omega^{vc}_{\bf k}
{a^v_{\bf k}}^\dagger a^c_{\bf k},
\ee
where we have defined  $\hbar \Omega^{\lambda \lambda'}_{\bf k} = {\bf M}^{\lambda \lambda'}
({\bf k}) \cdot {\bf A}$ to be the Rabi frequencies. 

Let us now consider an arbitrary two band system, whose low-energy quasiparticle bands are described by the generic $2 \times 2$ Hamiltonian,
\be\label{H_gen} 
 H_0 = \sum_{\bf k}{\bf h}_{\bf k} \cdot \bsigma,
\ee
 where $ {\bf h}_{\bf k} = (h_{0{\bf k}},h_{1{\bf k}}, h_{2{\bf k}}, h_{3{\bf k}})$ is a vector composed of real scalar elements and $ \bsigma = (\openone_2, \sigma_x, \sigma_y, \sigma_z)$ is a vector composed of the identity and the Pauli matrices in two dimensions.  
 The eigen energies of $H_0$ are given by $\varepsilon_{\bf k}^{\lambda} = h_{0{\bf k}} + \lambda g_{\bf k} $ where we have defined  $g_{\bf k} \equiv \sqrt{h_{1{\bf k}}^2+h_{2{\bf k}}^2+h_{3{\bf k}}^2} $, and $\lambda =1$ (or  $-1$) denotes the conduction (valance) band. 
 The corresponding  eigenvectors can be conveniently expressed as $\psi^{\lambda} = \{ \cos \theta_{\lambda \bf k}, \sin\theta_{\lambda \bf k} e^{i \phi_{\bf k}}\}$, where $\tan\phi_{\bf k} = h_{2 \bf k}/h_{1 \bf k}$, and $\tan\theta_{\lambda \bf k} =(\lambda g_{\bf k}-h_{3 \bf k}) / \sqrt{h_{1\rm k}^2 + h_{2 \rm k}^2}$. Note that for the special case of materials, such as graphene,  for which $h_{0 \bf k} = h_{3 \bf k} = 0$, we have $\cos\theta_{\lambda \bf k }  = 1/ \sqrt{2}$, and $\sin\theta_{\lambda \bf k } = \lambda/\sqrt{2}$. 
 
The optical matrix elements can accordingly be obtained in a very general form, as follows
\bearr \label{eq:Mvv}
{\bf M}^{vv} &=& \frac{e}{\hbar g_{\bf k}} \Big( g_{\bf k}\nabla_{\bf k} h_{0 {\bf k}}  - \sum_{i=1,2,3} h_{i\bf k}\nabla_{\bf k} h_{i{\bf k}}~\Big),  \\ 
{\bf M}^{cc} &=& \frac{e}{\hbar g_{\bf k}} \Big( g_{\bf k}\nabla_{\bf k} h_{0 {\bf k}}  + \sum_{i=1,2,3} h_{i\bf k}\nabla_{\bf k} h_{i{\bf k}}~\Big), \\ 
{\bf M}^{vc} &=& -\frac{e}{\hbar g_{\bf k} h_{\bf k}}\Big(- h_{\bf k}^2\nabla_{\bf k} h_{3 {\bf k}}  + (h_{1 {\bf k}} h_{3 {\bf k}} - i h_{2 {\bf k}} g_{{\bf k}})  \nabla_{\bf k} h_{1 {\bf k}} \nn \\
& & +  (h_{2 {\bf k}} h_{3 {\bf k}} + i h_{1 {\bf k}} g_{{\bf k}})  \nabla_{\bf k} h_{2 {\bf k}}  ~\Big),   \label{eq:Mcv}\\
{\bf M}^{cv}  &=&  \left({\bf M}^{vc}\right)^* 
\eearr
where we have defined $h_{\bf k}^2 \equiv h_{1{\bf k}}^2 + h_{2{\bf k}}^2$. 

To describe the dynamics of the system, we consider the time evolution of the momentum resolved density matrix ($2 \times 2$), whose diagonal elements are  
$\rho_{11} = \rho^v_{\bf k}$, and $\rho_{22} = \rho^c_{\bf k}$. Here $\rho_{\bf k}^\lambda \equiv \langle {a_{\bf k}^{\lambda}}^\dagger a_{\bf k}^{\lambda} \rangle$ denotes the  momentum resolved electron density in the valance and conduction bands. The off-diagonal elements of the 
density matrix are given by $\rho_{12} = p_{\bf k} \equiv \langle {a_{\bf k}^{c}}^\dagger a_{\bf k}^v \rangle$, and $\rho_{21} =  p_{\bf k}^*$, 
with $p_{\bf k}$ denoting the inter-band coherence or polarization. Using the equation of motion, $i \hbar \partial_t {\hat\rho}(t) = [H, \hat\rho]$, we obtain the following equation 
for the population inversion, $n_{\bf k} \equiv \rho^{c}_{\bf k} - \rho^{v}_{\bf k}$, as 
\be\label{bloch1}
\partial_t{n_{\bf k}} = 4  \Im  \left[\{{\Omega_{\bf k}^{vc}}(t)\}^* p_{\bf k}(t)\right]~.
\ee
The corresponding inter-band coherence evolves as
\be\label{bloch2}
\partial_t{p_{\bf k}} = i\left[ \omega_{\bf k}+ \Omega_{\bf k}^{cc}(t) -\Omega_{\bf k}^{vv} (t)\right] p_{\bf k}(t)-i\Omega_{\bf k}^{vc}n_{\bf k}(t)~,
\ee
where $ \hbar \omega_{\bf k} = \varepsilon_{\bf k}^c - \varepsilon_{\bf k}^v$.

To obtain the steady state (long time average response), we do a rotating wave approximation, where in we get rid of the fast oscillating terms (with frequencies $2 \omega$ and higher), and retain the slow time dependence in Eqs.~\eqref{bloch1} and \eqref{bloch2}. 
To this end, one can substitute ${\bf A}(t) = {\bf e_0} \omega^{-1} E_0  \cos{\omega t} $, with $\bf{e_0}$ denoting the polarization direction. Keeping only the low frequency resonant terms (of the form of $e^{\pm i(\omega - \omega_{\bf k})}$), while neglecting all the high frequency ones ($ e^{\pm i(\omega +\omega_{\bf k})}$) 
leads to:
\bearr \label{COBE2a}
\partial_t \tilde{n} & =& 2 \Im \left[ \tilde{\Omega}^{vc*}_{\bf k} \tilde{p}_{\bf k} \right]~, \\ 
\label{COBE2b}
\partial_{t}\tilde{p}_{\bf k} & = & i (\omega_{\bf k}-\omega) \tilde{p}_{\bf k}  - i \tilde{\Omega}^{vc}_{\bf k}
\tilde{n}_{\bf k}/2~,
\eearr
where $\hbar \tilde{\Omega} = {\bf M} \cdot {\bf e_0} ~E_0/\omega$,  $\tilde{p}_{\bf k}(t) = {p}_{\bf k}(t)e^{-i \omega t}$, and $ \tilde{n}_{\bf k} =n_{\bf k}(t) $. Here $\tilde{p}, ~{\rm and}~ \tilde{n}_{\bf k}$, are almost frozen (vary very slowly) over the timescales of the order of $1/\omega$.
Note that Eqs.~\eqref{COBE2a}-\eqref{COBE2b} do not include energy relaxation and decoherence mechanisms, arising due to electron-electron interactions,  electron-phonon, electron-impurity scattering and other interactions. To include these effects, phenomenological damping terms are added in the above equations [\onlinecite{meierbook}] leading to 
\bearr \label{COBE3a}
\partial_t \tilde{n}_{\bf k} & =& 2 \Im \left[ \tilde{\Omega}^{vc*}_{\bf k} \tilde{p}_{\bf k} \right] - \gamma_{1}(\tilde{n}_{\bf k}-n^{\rm eq}_{\bf k})~, \\ 
\label{COBE3b}
\partial_{t}\tilde{p}_{\bf k} & = & i (\omega_{\bf k}-\omega) \tilde{p}_{\bf k}  - i \tilde{\Omega}^{vc}_{\bf k} 
\tilde{n}_{\bf k}/2-\gamma_{2}\tilde{p}_{\bf k}~.
\eearr
Here $\gamma_{1}$ is the inverse of the relaxation time for the momentum resolved occupation 
number and 
$\gamma_{2}$ is the inverse relaxation time  of coherence. The equilibrium population inversion in absence of the  optical field is $n^{\rm eq}_{\bf k} = f_{c\bf k} - f_{v\bf k}$. The function $f_{a{\bf k}} = [1+ \exp{((\epsilon_{\bf k}^a-\mu)/k_B T)}]^{-1}$ denotes the Fermi function ($k_B$ and $T$ are the Boltzmann constant and temperature, respectively). The relaxation rates $\gamma_1$ and $\gamma_2$ are usually dominated by  electron-phonon and electron-electron interactions, respectively. Generally the damping rates are frequency dependent [\onlinecite{Mishchenko}], and can be modeled microscopically by  
self-consistently solving electron-electron (generally at a mean field level), electron-phonon and electron-photon coupling equations [\onlinecite{Andreas1}]. For example in graphene, typical values of $\gamma_{1} \approx 10^{12}$ Hz, and $\gamma_{2} \approx 10^{14}$ Hz for optical frequencies [\onlinecite{Zhang1}]. However, for simplicity of analysis, we choose these rates to be constants over the frequency range considered in this work. 

In the steady state regime, we can solve Eqs.~\eqref{COBE3a} and \eqref{COBE3b} to obtain the following steady state values for the population inversion
\be\label{n0}
\frac{{\tilde n}_{\bf k}}{n^{\rm eq}_{\bf k}}  = \left(1 + \frac{\gamma_{2}| \tilde{\Omega}^{vc}|^2}{ \gamma_{1} [(\omega_{\bf k}-\omega)^2 + \gamma_{2}^2] } \right)^{-1}~,
\ee
and the inter-band coherence
\be\label{p0}
\tilde{p}_{\bf k}  = \frac{\tilde{n}_{\bf k}}{2} \frac{\tilde{\Omega}^{vc}}{(\omega_{\bf k}-\omega) + i \gamma_{2}}~.
\ee
Before proceeding further it is instructive to express the steady state population inversion of Eq.~\eqref{n0}, in the following form,
\be \label{Gdef}
 \frac{{\tilde n}_{\bf k}}{n^{\rm eq}_{\bf k}}  \equiv G= \left(1 +  \frac{|{\bf M}^{vc} \cdot {\bf e_0}|^2}{e^2 v_{\rm F}^2} \zeta^2 \frac{ \gamma_{2}^2}{ [(\omega_{\bf k}-\omega)^2 + \gamma_{2}^2] } \right)^{-1}~,
\ee
where $\zeta$ is specified by Eq.~\eqref{zeta}. In Eq.~\eqref{Gdef}, ${\bf M}^{vc} \cdot {\bf e_0}/(e v_{\rm F})$ is the dimensionless material dependent optical matrix element component which couples to the incident radiation. The term $\gamma_2/{[(\omega_{\bf k}-\omega)^2 + \gamma_{2}^2]}$ in Eq.~\eqref{Gdef}  is a Lorentzian centered around $\omega = \omega_{\bf k}$ with half-width $\gamma_{2}$. It reduces to a Dirac-delta function 
$\delta(\omega-\omega_{\bf k})$ in the limiting case of $\gamma_{2}/\omega \to 0$. Furthermore, we note that $G$ can be expanded in a power series of $\zeta^2$. Thus $\zeta$ is a 
dimensionless effective field strength, which, for a fixed value of the optical frequency and the damping constants, distinguishes between the linear ($\zeta \ll 1$) and the nonlinear ($\zeta \geq 1$) response regime [\onlinecite{Mishchenko}]. Additionally, for a fixed value of the optical frequency and effective field strength, the ratio $\gamma_{2}/\omega$ determines the clean/high frequency ($\gamma_{2} \ll \omega$) or the dirty/low frequency limits ($\gamma_{2} \ge \omega$).  This  classification of the system's response into linear and non-linear regimes, or alternatively into the clean and the dirty limit arises naturally in our formulation and will be used in the rest of the manuscript. 

\begin{figure}[t!]
\centering
\includegraphics[width=1.0\linewidth]{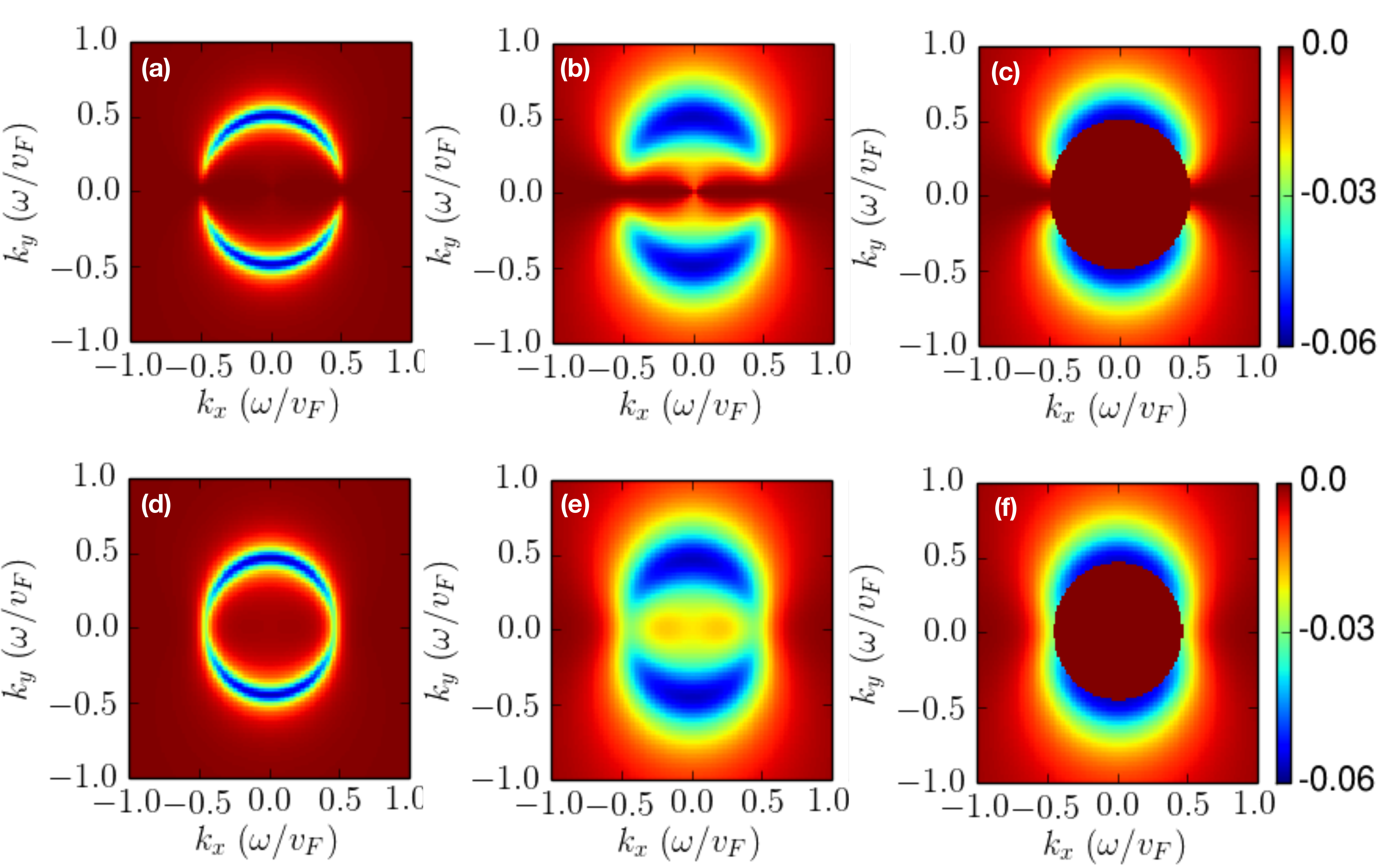}
\caption{ The momentum resolved current $\tilde{J}_{{\bf k}x} $ density of Eq.~\eqref{eq:J012}, in the $k_x-k_y$ plane for graphene in panels a), b) and c) and for massive graphene in panels d), e), and f). In panels b) and c) [panels  e)-f) for massive case] we have a larger coherence decay rate $\gamma_2 = \omega/5$ as compared to a) for which $\gamma_2 = \omega/20$ and this show the impact of broadening  of the current density around the circle $k = \omega/(2 v_{\rm F})$ at $\mu = 0$. Further in panel c) for graphene [panel f) for massive graphene], we display the impact of a finite chemical potential $\mu = 0.5 \hbar \omega$ which manifests itself in the Pauli blocking of the momentum space for $k < \mu/\hbar v_{\rm F}$. For more details see Eq.~\eqref{eq:J02} for massless graphene, and Eq.~\eqref{eq:DeltaJ} for the massive graphene case. 
\label{fig2}}
\end{figure}

Having obtained the steady state density matrix elements, we now proceed to calculate the induced current density and the nonlinear optical conductivity. 

\section{Nonlinear inter-band optical conductivity}
\label{sec3} 
To obtain the inter-band contribution to the optical current and optical conductivity, let us consider the generic form of the current density for a $d$-dimensional system:   ${\bf J}(t) = -g_s g_v (2 \pi)^{-d} 
\int d{\bf k} ~{\bf J}_{\bf k}(t)$, where  $g_s$ ($g_v$) denotes the spin (valley) degeneracy factor, ${\bf J}_{\bf k}(t)  = 
e {\rm Tr}[\rho_{\bf k}(t) {\bf v}_{\bf k}(t) ]$ and the generalized velocity operator is given by  $\hat{\bf v}_{\bf k} = -i \hbar^{-1} [{\bf r}, H_0] = \hbar^{-1} \nabla_{\bf k} H_0$. 
The momentum dependent component of the particle current density is then 
\be \label{eq:J0}
{\bf J}_{\bf k}(t) =  2 \Re e [p_{\bf k} {\bf M}^{cv}_{\bf k}] + \sum_{\lambda = c,v}\rho_{\bf k}^\lambda  {\bf M}^{\lambda \lambda}~,
\ee
In presence of particle-hole symmetry, as in graphene or gapped graphene, we have ${\bf M}_{\bf k}^{cc} = -{\bf M}_{\bf k}^{vv}$. One can then rewrite Eq.~\eqref{eq:J0} as 
 \be \label{eq:J01}
 {\bf J}_{\bf k}(t) =  2 \Re e [p_{\bf k}(t) {\bf M}^{cv}_{\bf k}] + n_{\bf k}(t) {\bf M}^{cc}_{\bf k}~. 
 \ee
It is evident that the first term in Eq.~\eqref{eq:J01} arises from the inter-band contribution, while the second term originates from the 
intra-band contributions. In Eq.~\eqref{eq:J01}, $p_{\bf k} (t) = \tilde{p}_{\bf k} e^{i \omega t}$, and consequently 
the momentum resolved current density consists of three terms: $2\Re e\left[\tilde p_{\bf k}{\bf M}_{\bf k}^{cv}\right]\cos{\omega t}$,
$-2\Im m \left[\tilde p_{\bf k}{\bf M}_{\bf k}^{cv}\right]\sin{\omega t}$ and $\tilde n_{\bf k}{\bf M}_{\bf k}^{cc}$. Of these, the term proportional to $\cos{\omega t}$ is the out of phase (with respect to the incident field) response of the system and it does not contribute to the dissipative part of the conductivity [\onlinecite{Mishchenko}]. Hence it will be neglected in the rest of the article. 
Furthermore, for graphene and gapped graphene, it is easy to check that the intra-band term $\tilde n_{\bf k}{\bf M}_{\bf k}^{cc}$, vanishes on performing the momentum sum. In fact, since we have considered only momentum conserving vertical transitions in calculating the electronic density matrix, it then follows that the intra-band part of Eq.~\eqref{eq:J0} should vanish for all materials after the ${\bf k}-$integration. 

One therefore needs to focus only on the dissipative part of the current, which is captured by the in-phase (to the electric field) part of the response corresponding to the $\sin(\omega t)$ term. 
This part of the momentum resolved current, can be expressed as ${\bf J}_{\bf k} (t) = {\bf {\tilde J}}_{\bf k} \sin(\omega t) $, with
 \be \label{eq:J012}
 {\bf {\tilde J}}_{\bf k} = - \frac{E_0}{\hbar \omega}{\tilde n}_{\bf k} \Im m \left\lbrace\frac{({\bf M}^{vc}_{\bf k}\cdot {\bf e_0}){\bf M}_{\bf k}^{cv}}{\omega_{\bf k}-\omega+i \gamma_{2 }} \right\rbrace ~.
 \ee
The corresponding dynamical nonlinear optical conductivity can then be obtained by integrating the above expression and using $\sigma_{ij} = |\tilde J_i|/[E_0 ({\bf e_0}\cdot \hat{j})]$,  where $E_0 ({\bf e_0}\cdot \hat{j})$  is amplitude  of the field along the $\hat{j}$ direction.

If one considers a linearly polarized light, say along the $x$-direction, and restrict oneself to the longitudinal response only,  then Eq.~\eqref{eq:J012} reduces to
\be \label{eq:J013}
\frac{{{\tilde J}}_{{\bf k}x}}{E_0} = - \frac{{\tilde n}_{\bf k} |M^{vc}_x|^2}{\hbar \omega} \Im m \left\lbrace\frac{1}{\omega_{\bf k}-\omega+i \gamma_{2}} \right\rbrace ~.
\ee
In Eq.~\eqref{eq:J013}, the total current, is in general `cut-off' or band-width dependent. For systems described by a continuum model, the ultraviolet momentum cutoff is inversely proportional to the lattice spacing. On the contrary, for systems described by a tight-binding model, the energy cutoff is typically the band-width of the system. 
Furthermore, there is static component of the current in the $\omega \to 0$  limit. This is unphysical, since in this limit, the inter-band current must vanish for a time-independent vector potential. Accordingly, this contribution needs to be subtracted from Eq.~\eqref{eq:J013} as prescribed in Ref.~[\onlinecite{Falkovsky2007}]. 

The momentum resolved current density of Eq.~\eqref{eq:J013} is shown in Fig.~\ref{fig2} for graphene and massive graphene.  

Using Eq.~\eqref{eq:J013}, the longitudinal optical conductivity in the most general {\it nonlinear-dirty} case can be expressed as 
 \be \label{eq:cond}
 \sigma_{xx}(\omega) = \frac{g_s g_v}{\hbar \omega (2 \pi)^d} \int d{\bf k} |M^{vc}_x|^2 {\tilde n}_{\bf k} \Im m \left\lbrace\frac{1}{\omega_{\bf k}-\omega+i \gamma_{2}} \right\rbrace.
 \ee
We emphasize here that the non-linearity of the optical conductivity in Eq.~\eqref{eq:cond}, stems from the $n_{\bf k}$ term, whose solution is obtained from the optical Bloch equation within RWA.  

Let us now consider the following limiting cases in which Eq.~\eqref{eq:cond} simplifies. The limiting case for $\gamma_2/\omega  \ll 1$ ($\gamma_2/\omega  \ge 1$) corresponds to the clean (dirty) limit, and the limiting case for $\zeta \ll 1$ ($\zeta \ge 1$) is related to the linear (non-linear) response of the system.

\subsection{Linear response in the clean limit: $\zeta \ll 1$, and $\gamma_{2}/\omega \ll 1$}

In this limit, to zeroth order in $\zeta$, we have $n_{\bf k} \to {n_{\bf k}^{\rm eq}}$, and   
converting the Lorentzian into a Dirac-delta function,  we have 
\be \label{eq:aa}
 \sigma_{ xx}^{\rm lc}(\omega)  =  \frac{-\pi g_s g_v}{\hbar \omega (2 \pi)^d} \int d{\bf k} |M^{vc}_x|^2 \delta(\omega_{\bf k}-\omega) (f_{c\bf k}-f_{v \bf k})~.
\ee
Note that the conductivity obtained above [Eq.~\eqref{eq:aa}], is identical to that obtained from the Kubo formalism. 

For systems with particle-hole symmetry and isotropic quasi-particle dispersion, such as graphene and massive graphene,  Eq.~\eqref{eq:aa} can be expressed as
\be \label{eq:ab}
 \sigma_{xx}^{\rm lc}(\omega) =  \frac{\pi g_s g_v g(\omega, \alpha, T)}{\hbar \omega (2 \pi)^d} \int d{\bf k} |M^{vc}_x|^2 \delta(\omega_{\bf k}-\omega) ~,
 \ee
where $\alpha \equiv {\rm max}\{\mu, \Delta\}$, with $\Delta$ being half of the band-gap in a given semiconductor.  The function
 \be \label{eq:g}
g(\omega, \alpha, T) \equiv \frac{1}{2}\left[
\tanh\left(\frac{\hbar \omega + 2 \alpha}{4 k_B T}\right) + \tanh\left(\frac{\hbar \omega - 2 \alpha}{4 k_B T}\right)\right]~.
\ee
In the zero temperature limit, $g(\omega, \alpha, T\to 0) = \Theta( \hbar \omega/2 - |\alpha|)$, where $\Theta(x)$ denotes the Heaviside step function. 

\subsection{Linear response in the dirty limit: $\zeta \ll 1$, and $\gamma_{2 }/\omega \ge 1$}

As in the previous case, here again we can approximate $n_{\bf k} \to {n_{\bf k}^{\rm eq}}$ upto zeroth order in $\zeta$. However in this case the Lorentzian has to be retained in the integral. The corresponding conductivity is given as  
\be \label{eq:bb}
 \sigma_{xx}^{\rm ld}(\omega) = \frac{- g_s g_v}{\hbar \omega (2 \pi)^d} \int d{\bf k} |M^{vc}_x|^2  \frac{\gamma_{2} (f_{c\bf k}-f_{v \bf k})}{(\omega_{\bf k}-\omega)^2+ \gamma_{2}^2}~.
 \ee

\subsection{Non-linear response in the clean limit: $\zeta \ge 1$, and $\gamma_{2 }/\omega \ll 1$ } 
Here the Lorentzian can be approximated by a delta function, and from Eq.~\eqref{eq:cond} one obtains 
 \be \label{eq:con}
 \sigma_{xx}^{\rm nc}(\omega) = \frac{-\pi g_s g_v}{\hbar \omega (2 \pi)^d} \int d{\bf k} |M^{vc}_x|^2 \frac{\delta(\omega_{\bf k}-\omega) (f_{c \bf k}-f_{v \bf k})}
 {1+ |{M}^{vc}_{x}|^2 \zeta^2/(e^2 v_{\rm F}^2)}~.
 \ee
For systems with particle-hole symmetry and isotropic quasi-particle dispersion, such as graphene and massive graphene, Eq.~\eqref{eq:aa} can be approximated as
\be\label{eq:cc}
 \sigma_{xx}^{\rm nc}=  \frac{\pi  g_s g_v g(\omega, \alpha, T)}{\hbar \omega (2 \pi)^d} \int d{\bf k} |M^{vc}_x|^2 \frac{\delta(\omega_{\bf k}-\omega)}
 {1+ \zeta^2 |{M}^{vc}_{x}|^2 /(e^2 v_{\rm F}^2)}~,
 \ee
where $g(\omega, \alpha, T)$ is defined in Eq.~\eqref{eq:g}.

Below, we will explore the optical conductivity of doped and gapped graphene in all the four regimes. Note that the transverse (Hall like) conductivity, $\sigma_{yx}(\omega)$ will also have expressions similar to that of Eqs.~\eqref{eq:cond}-\eqref{eq:bb}, with the substitution $|M^{vc}_x|^2 \to M^{vc}_x M^{cv}_y $ in the numerator.

Here we would like to emphasize that the formalism described above gives only the finite frequency  part of the conductivity which cannot be extrapolated to the DC limit (limitation imposed due to RWA).  The full paramagnetic conductivity, which includes the zero frequency Drude weight ($\cal D$) is given by $ \sigma_{\rm total}(\omega) = \pi {\cal D} \delta(\omega) + \sigma(\omega)$. The Drude weight is explicitly given by ${\cal D} = \lim_{\omega \to 0} \omega \Im m[\sigma(\omega)]$ (see Ref.~[\onlinecite{Stauber_NJP}] for details). The imaginary part of the optical conductivity can therefore be evaluated by making use of the Kramers-Kronig relations, as  
\be
\Im m \sigma(\omega ) = \frac{2}{\pi \omega} {\cal P} \int_0^\infty d \omega' \frac{\omega'^2 \Re e \sigma(\omega')}{\omega^2-\omega'^2}~, 
\ee
where ${\cal  P}$ denotes the principal part of the integral. 

Experimentally the optical conductivity is probed via the differential optical transmission or reflection spectroscopy. Motivated by this, in the next Section, we  explore the effect of non-linearity on the reflection and transmission spectrum.

\section{Non-linear differential transmissivity and reflectivity for 2D materials}
\label{sec4}
The calculated non-linear conductivity has significant implications for optical experiments.  
In this section we determine the differential transmission (or equivalently the reflection coefficient) of a `free standing' two dimensional material, suspended in vacuum. 

Let us consider a linearly polarized incident field propagating normal to the plane of the 2D material. 
For definiteness, let us assume that the two-dimensional sample is freely suspended 
in the $x-y$ plane at $z = 0$. An incoming field directed along the negative
$\hat{z}$-direction can be decomposed
into its components as,
\be \label{Ecomp}
{\bf E} =
\left\{
        \begin{array}{ll}
                 {\bf E}_I\sin(\omega t + qz) +  {\bf E}_R\sin(\omega t - qz) & , z > 0 \\
                 {\bf E}_T\sin(\omega t + qz) & , z < 0,
        \end{array}
\right.
\ee
where $q = \omega/c$ is the photon momentum and $ {\bf E}_i = (E_{0i}\cos\theta_0, E_{0i}\sin\theta_0)$,  with $ i = I,
R~\text{and}~T$, denoting the incident, reflected and the transmitted components.  
$\theta_0$ is the polarization angle with respect to the $\hat{x}$-axis. 
Interaction of the incident light with the carriers in the 2D sample, is modeled in the Maxwell's equations, via the total current density 
in the 2D ($\hat{x}-\hat{y}$) plane, having a vanishing thickness. This implies  ${\bf J}(t) \equiv
{\bf J}(t)\delta(z)$. The spatiotemporal evolution of the $\hat {x}$ component of
the electromagnetic field is given by 
\be \label{eq:Ex}
{\bf\nabla}^2E_x -\mu_0\varepsilon_0\frac{\partial^2E_x}{{\partial t}^2} =
\mu_0\partial_t J_x\delta(z)~.
\ee
Integrating the above equation across the two dimensional plane yields
\be \label{eq:Ex2}
\partial_zE_x|_{z=0^{+}} - \partial_zE_x|_{z=0^{-}} = \mu_0\partial_tJ_x~.
\ee
Substituting Eq.~\eqref{Ecomp} in Eq.~\eqref{eq:Ex2}, we obtain 
\be
(E_{I}-E_{R}-E_{T})\cos\theta_0 = \frac{ c \mu_0}{ \omega \cos(\omega t)}
\partial_tJ_x~.
\ee
Additionally, the continuity of the electromagnetic field across the 2D layer yields,
\be
E_{I}+E_{R} = E_{T}.
\ee
Furthermore, if we assume the material to be isotropic, with a conductivity tensor which has only diagonal elements, we have $J_x = \sigma_{xx}  \sin(\omega t) [\omega  E_T \cos(\theta_0)]$.  With this the non-linear transmissivity can be easily obtained to be, 
\be
T(\omega) \equiv   \left| \frac{E_{T}}{E_{I}} \right|^2 =  \left[1 +  \frac{\pi \alpha_{\rm fine}}{2} \frac{\sigma_{xx}(\omega)}{\sigma_0} \right]^{-2} ~, 
\ee
 where $\alpha_{\rm fine} \equiv e^2/(4 \pi \hbar c \epsilon_0)$ is the universal fine structure constant and $\sigma_0 \equiv e^2/(4 \hbar)$, is the so called universal ac conductivity of graphene.
Here $\sigma_{xx}(\omega)$ is the non-linear longitudinal conductivity.  The reflectivity $R(\omega) = |E_R/E_I|^2$, can also be obtained in a similar fashion and it is given as 
\be
R(\omega) = \left(\frac{\pi \alpha_{\rm fine}}{2} \frac{\sigma_{xx}(\omega)}{\sigma_0}\right)^2\left[1+  \frac{\pi \alpha_{\rm fine}}{2} \frac{\sigma_{xx}(\omega)}{\sigma_0} \right]^{-2}~.
\ee
The absorption coefficient is given by $\alpha(\omega) \equiv 1- T(\omega) - R(\omega)$ and it denotes the fraction of light intensity  which is either scattered by the surface atoms of the 2D material (Rayleigh scattering) or absorbed. 

\section{Nonlinear conductivity of doped graphene}
\label{sec5}

\begin{figure}[t!]
\centering
\includegraphics[width=0.99\linewidth]{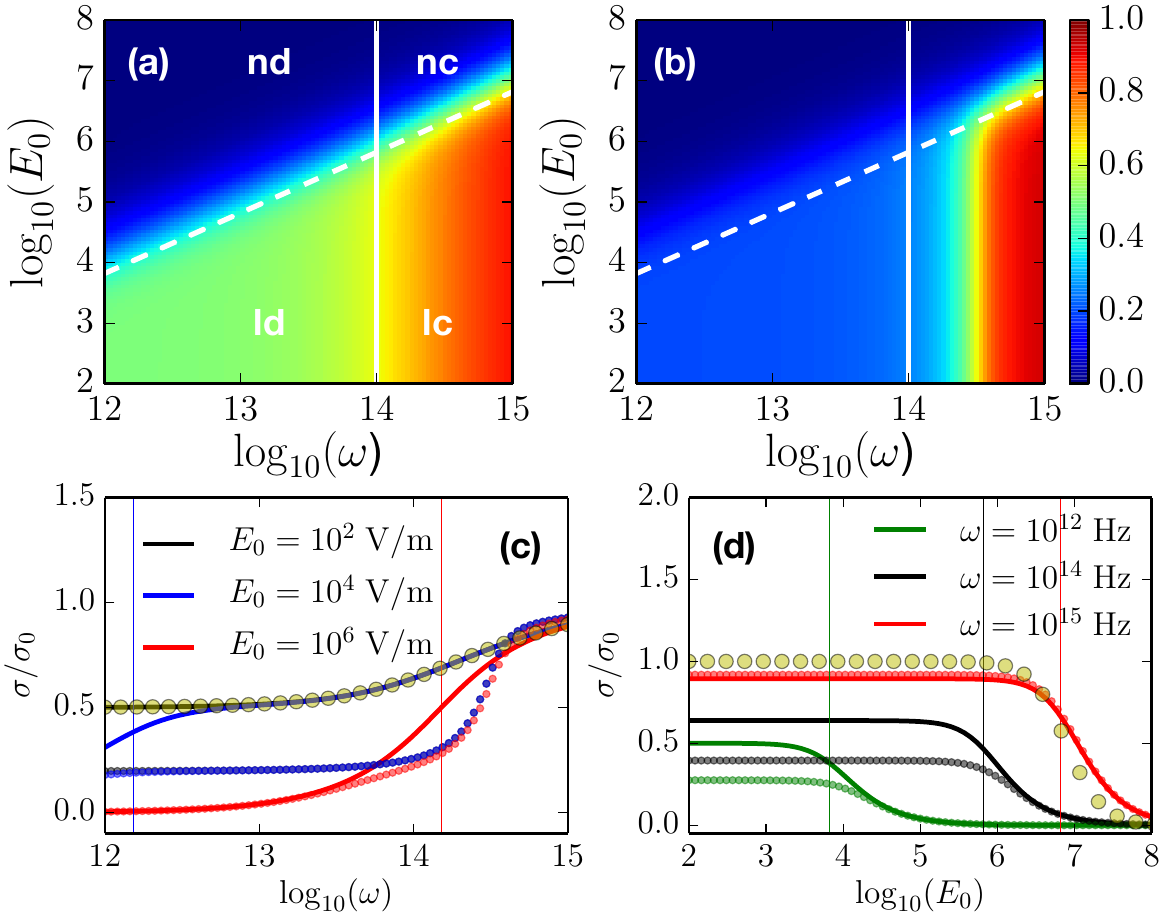}
\caption{Color plot of the nonlinear (inter-band) optical conductivity (in units of $\sigma_0 = e^2/4 \hbar$) as a function of frequency and the electric field strength of the incident laser beam for (a) pristine graphene with $\mu =0$  and (b) doped graphene with  $\mu = 0.1 $ eV (or equivalently $\mu/\hbar = 1.5 \times 10^{14}$ Hz). The vertical (white) line represents $\omega = \gamma_2$, which is the boundary of the clean (also called collision-less or high frequency limit for $\omega \ge \gamma_2$) and the dirty (or collisional or low frequency) limit.
The dashed white line marks the boundary of the linear-nonlinear response regime [$\zeta \equiv e v_{\rm F} E_0/(\hbar \omega \sqrt{\gamma_1 \gamma_2}) = 1$], and together with the boundary of the clean limit, it divides the plot into 4 regimes, linear clean (marked `lc'), non-linear clean (`nc'), non-linear dirty (`nd') and linear dirty (`ld'). Panel (c) shows horizontal cuts from the upper two panels, {\it i.e.}, the conductivity as a function of $\omega$ for different electric field strengths for both, the $\mu =0$ case (solid lines), and the doped case of $\mu = 0.1$ eV (dotted lines of the same color). The yellow circles show the excellent match of Eq.~\eqref{eq:ldgr1}, in the linear dirty limit, with the exact numerical results.
Panel (d) displays vertical cuts from the upper two panels, {\it i.e.}, the conductivity as a function of $E_{0}$ for different frequencies with solid lines for the $\mu =0$, and the dotted lines of the same color for $\mu = 0.1 eV$. Here the yellow circles show the excellent match of Eq.~\eqref{eq:cd}, in the non-linear clean limit, with the exact numerical results. In all the panels we have chosen $v_{\rm F} =10^6$ m/s, and $\{\gamma_1,\gamma_2\} = \{10^{12},10^{14}\}$ Hz.
\label{fig4}}
\end{figure}

\begin{figure}[t!]
\centering
\includegraphics[width=0.99\linewidth]{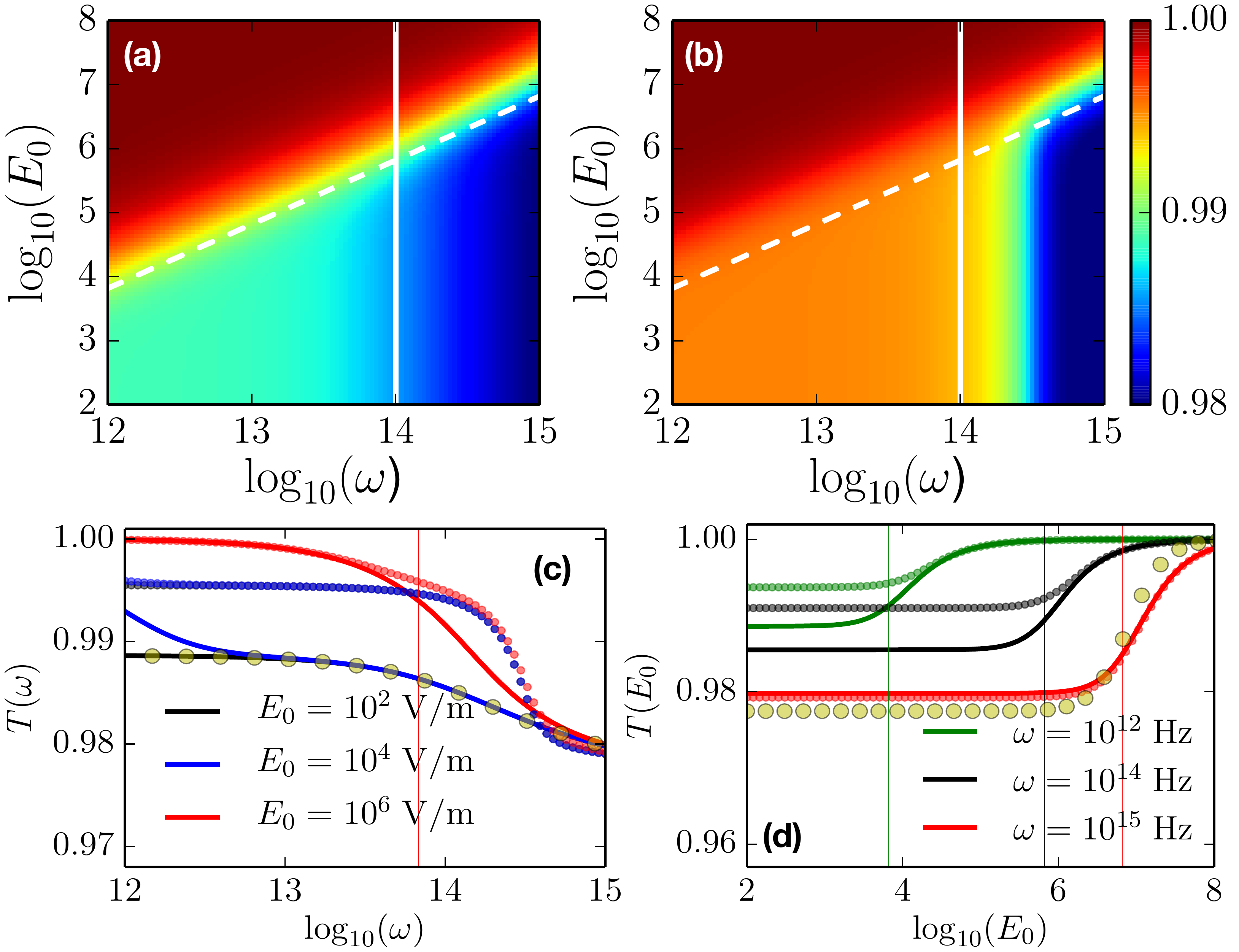}
\caption{Color plot of the nonlinear transmission as a function of frequency and the electric field strength of the incident laser beam for (a) pristine graphene with $\mu =0$  and (b) doped graphene with  $\mu = 0.1 $ eV (or equivalently $\mu/\hbar = 1.5 \times 10^{14}$ Hz). Panels (c) and (d) display horizontal and vertical cuts, respectively with the solid lines corresponding to the $\mu = 0$ case, and the dotted lines representing the $\mu = 0.1$ eV. 
Other parameters are identical to that of Fig.~\ref{fig4}.
\label{fig3b}}
\end{figure}

In this section we apply the general framework developed in Sec.~\ref{sec2} and Sec.~\ref{sec3} to calculate the non-linear optical conductivity of graphene. For simplicity, we use the effective low energy quasiparticle dispersion of graphene instead of the full tight-binding Hamiltonian. The corrections to the low energy dispersion on the universal conductivity has been shown to be $ \sim \hbar\omega/(72 \times 2.8 {\rm eV})$ [\onlinecite{Stauber_TB}]. Thus 
such effects can be safely neglected for all frequencies in the optical domain and below. 
For graphene, we have $\varepsilon_{\bf k}^c = -\varepsilon_{\bf k}^v = \hbar v_{\rm F} k$, and 
${\bf M}^{vc} = {\bf M}^{cv*} = iev_{\rm F} \tau \{\sin{\phi_{\bf k}}, -\cos{\phi_{\bf k}}\}$ where  
 $k = ({k_x^2 + k_y^2})^{1/2}$,  $\phi_{\bf k} = \tan^{-1}(k_y/k_x)$ and $\tau = +1$ ($-1$) for the $K$ ($K'$) valley. 

We now consider the non-linear optical conductivity of graphene arising from the inter-band transitions in various regimes. We  start with  the {\it linear response in the clean limit} of graphene, {\it i.e.,} for $\zeta \ll 1$ and $\gamma_2/\omega \ll 1$ limit. 
In this regime,  using Eq.~\eqref{eq:ab}, one can obtain the finite temperature optical conductivity for graphene as 
\be \label{eq:abgr}
\sigma_{xx}^{\rm lc}(\omega) = \frac{g_sg_ve^2}{16\hbar} g(\omega, \mu, T)~. 
\ee
This has also been derived earlier from the Kubo formalism (see Eq.~(25) of Ref.~[\onlinecite{Stauber_TB}]). In Eq.~\eqref{eq:abgr},
$g_s =2$ ($g_v =2$) denotes the spin (valley) degree of freedom in graphene and the function $g(\omega, \mu, T)$ is defined in Eq.~\eqref{eq:g}. In the limiting case of $T \to 0$, including the spin and valley degeneracy factors, Eq.~\eqref{eq:abgr} reduces to, 
\be \label{eq:abgrT0}
\sigma_{xx}^{\rm lc}(\omega) = \frac{e^2}{4\hbar}\Theta\left(\frac{\hbar\omega}{2} - |\mu|\right),
\ee
which has a finite universal value as long as the optical excitation energy is greater than twice the chemical potential. 
Equation~\eqref{eq:abgrT0} yields the so called `universal' ac conductivity of graphene which was predicted in Refs.~[\onlinecite{Ando2002, Gusynin2006}], and experimentally observed in Ref.~[\onlinecite{Nair2008}]. 

In the {\it non-linear and clean} limit, we have $\zeta \ge 1$ and $\gamma_2/\omega \ll 1$. Using Eq.~\eqref{eq:cc}, the optical conductivity for graphene takes the form 
\be \label{eq:cd}
\sigma_{xx}^{\rm nc}(\omega) = \frac{e^2}{4\hbar}\frac{2}{\zeta^2}\left( 1 - \frac{1}{(1 + \zeta^2)^{1/2}}\right)  g(\omega, \mu, T)~.
\ee
As a consistency check we note that Eq.~\eqref{eq:cd} reduces to Eq.~\eqref{eq:abgr} in the limiting case of $\zeta \to 0$.
Note that for the specific case of pristine graphene ($\mu=0$ at zero temperature), Eq.~\eqref{eq:cd} can also be derived by following the wave-function based approach and integrating Eq.~(12) of Ref.~[\onlinecite{Mishchenko}]. 

Next we consider the {\it linear response in the dirty limit} for which $\zeta \ll 1$ and $\gamma_2/\omega \geq 1$.
In this case the form of conductivity changes substantially and at zero temperature it is 
\be \label{eq:ldgr}
\sigma_{xx}^{\rm ld}(\omega) = \frac{e^2\gamma_2}{4\pi\hbar\omega}\int_{\frac{2|\mu|}{\hbar}}^{\frac{2\Lambda}{\hbar}}d\omega_{\bf k}\left[\frac{\omega_{\bf k}}{(\omega_{\bf k}-\omega)^2 + \gamma_2^2}- (\omega \to 0) \right]~,
\ee
where $\Lambda$ is the ultraviolet energy cutoff, which physically should correspond to half of the bandwidth. In a tight-binding model of graphene half of the bandwidth is $3\times 2.8$ eV.

Evaluating Eq.~\eqref{eq:ldgr} we obtain
\be \label{eq:ldgr1}
\sigma_{xx}^{\rm ld}(\omega) = \frac{e^2}{4 \pi \hbar} \left[f\left(\omega,  2 \Lambda/\hbar \right) - f\left(\omega, 2|\mu|/\hbar \right) \right]~,  
\ee
where 
\be 
f(\omega, x) \equiv \tan^{-1}\left(\frac{x-\omega}{\gamma_2}\right)+\frac{\gamma_2}{2\omega}
\ln\left[\frac{\gamma_2^2 + (\omega - x)^2 }{\gamma_2^2 + x^2}\right]~.
\ee
The finite temperature generalization of Eq.~\eqref{eq:ldgr1} has to be calculated numerically. 
As a consistency check we note that Eq.~\eqref{eq:ldgr1} reduces to Eq.~\eqref{eq:abgrT0} in the limit $\gamma_2 \ll \omega$.

For the most general case of the {\it nonlinear-dirty} regime, starting with Eq.~\eqref{eq:J012}, we express the dissipative part of the steady state current for each spin and valley of graphene ($\hat{x}$ and $\hat{y}$ components) as 
${\bf \tilde J}_{\bf k} = { \tilde J}_{\bf k}\left(\sin{\phi_{\bf k}},-\cos{\phi_{\bf k}}\right)$~. Here, 
\be \label{eq:J02}
{ \tilde J}_{\bf k} = \frac{e^2v_{\rm F}^2E_0\sin(\phi_{\bf k}-\theta_0)\gamma_{1}\gamma_{2}n_{\bf k}^{\rm eq}}{\hbar \omega \left\lbrace\gamma_{1}\left[(\omega_{\bf k}-\omega)^2 + \gamma_{2}^2\right]
+ \gamma_{2}|\tilde\Omega^{vc}|^2\right\rbrace}~,
\ee
and $\theta_0$ is the polarization angle of the linearly polarized field, with respect to the $\hat{x}$ axis. We then have the total longitudinal conductivity for graphene as $\sigma_{xx}(\omega) =(2 \pi)^{-2} g_s g_v \int d{\bf k} \ \sigma'_{xx}({\bf k}) $, where 
\be \label{Eq:ndgr}
\omega \sigma'_{xx}({\bf k}) = \frac{e^2}{\hbar} \frac{v_{\rm F}^2 \gamma_{2} (f_{c \bf k} -f_{v \bf k}) \sin^2\phi_{\bf k} }{\left[(\omega_{\bf k}-\omega)^2 + \gamma_{2}^2 (1+  \zeta^2 \sin^2\phi_{\bf k})\right] } - (\omega \to 0).
\ee
At $T \to 0$, the $k$-integral of Eq.~\eqref{Eq:ndgr} has lower and upper limits of $2 |\mu|/\hbar v_{\rm F}$  and $2 \Lambda/ \hbar v_{\rm F}$ respectively. It can therefore be evaluated exactly (as in Eq.~\eqref{eq:ldgr}) to obtain, 
\be \label{nd:gr}
\sigma_{xx}(\omega) = \frac{e^2}{4 \pi^2 \hbar} \int_{0}^{2 \pi} d \phi_{\bf k}\sin^2\phi_{\bf k} \left[f_1(2 \Lambda/\hbar) - f_1(2 |\mu|/\hbar)\right]~,
\ee
where, 
\be 
f_1(\omega, x) \equiv \frac{\gamma_2}{\gamma_{\phi_{\bf k}}}\tan^{-1}\left(\frac{x-\omega}{\gamma_{\phi_{\bf k}}}\right)+\frac{\gamma_2}{2\omega}
\ln\left[\frac{\gamma_{\phi_{\bf k}}^2 + (\omega - x)^2 }{\gamma_{\phi_{\bf k}}^2 + x^2}\right]~,
\ee
and $\gamma_{\phi_{\bf k}} = \gamma_2 (1+ \zeta^2 \sin^2\phi_{\bf k} )^{1/2}$. The $\phi_{\bf k}$ integral in Eq.~\eqref{nd:gr}, can also be done analytically, yielding a cumbersome expression without much insight.  
However, it is easy to check that in the linear response regime, $\zeta \to 0$, $\gamma_{\phi_{\bf k}} \to \gamma_2$, and Eq.~\eqref{nd:gr}, reduces to Eq.~\eqref{eq:ldgr1}, as expected. 

The optical conductivity of graphene, in all four regimes is shown in Fig.~\ref{fig4}. The $\zeta = 1$ line clearly marks the boundary of the non-linear response regime, with saturation effects dominating on the $\zeta >1$ side. Interestingly enough, the `universal' optical conductivity of graphene $\sigma_0 =e^2/(4 \hbar)$, seems to be valid only in the $\omega > \gamma_2$ and $\zeta \ll 1$ regime. Figure \ref{fig4} also suggests that non-linear optical saturation effects will become dominant with decreasing optical frequencies while keeping the laser intensity constant. The effect of non-linear optical conductivity on the transmission spectrum 
is highlighted in Fig.~\ref{fig3b}.

An interesting observation from Eq.~\eqref{eq:J02}, is that the off-diagonal conductivity $\sigma_{yx} (\omega)$ (for a $\hat x$ polarized field) vanishes for graphene. This is on account of the $\phi_{\bf k}$ integration vanishing for each valley. This is not the case for massive graphene where $\sigma_{yx} (\omega)$ is finite for each valley with opposite signs for the two valleys. Therefore the total $\sigma_{yx} (\omega)$ cancels out. 
This implies that one can possibly have a finite $\sigma_{yx}$ in massive graphene if the two valleys can be made to have a different bandgap.

\section{Nonlinear optical conductivity of gapped graphene}
\label{sec6}
We now proceed to discuss the case of gapped graphene. 
If a gap, $\Delta$ is introduced in the band structure of graphene, say by growing it epitaxially on top of SiC [\onlinecite{gap1_gr, gap1_gr_2}],  then the effective Hamiltonian would take the form of Eq.~\eqref{H_gen}, with $ h_{0{\bf k}} = 0$, $h_{1{\bf k}} = \hbar v_{\rm F} k_x$, $h_{2{\bf k}} = \hbar v_{\rm F} k_y$,
and $ h_{3{\bf k}} = \Delta$. In this case, we have $\varepsilon_{\bf k}^c = -\varepsilon_{\bf k}^v \equiv g_{\bf k} = (\hbar^2 v_{\rm F}^2 k^2 + \Delta^2)^{1/2}$. 
The $\hat x$ and $\hat y$ components of the inter-band optical matrix element  are given by 
\be \label{eq:Md}
\frac{{\bf M}^{vc}}{ev_{\rm F}} = -\left(\frac{\Delta \cos\phi_{\bf k}}{g_{\bf k}}- i \tau \sin\phi_{\bf k}  , \frac{\Delta \sin\phi_{\bf k}}{g_{\bf k}} + i \tau \cos\phi_{\bf k} \right), 
\ee
where $\tau = +1$ ($-1$) for the $K$-valley ($K'$-valley).

\begin{figure}[t!]
\centering
\includegraphics[width=0.99\linewidth]{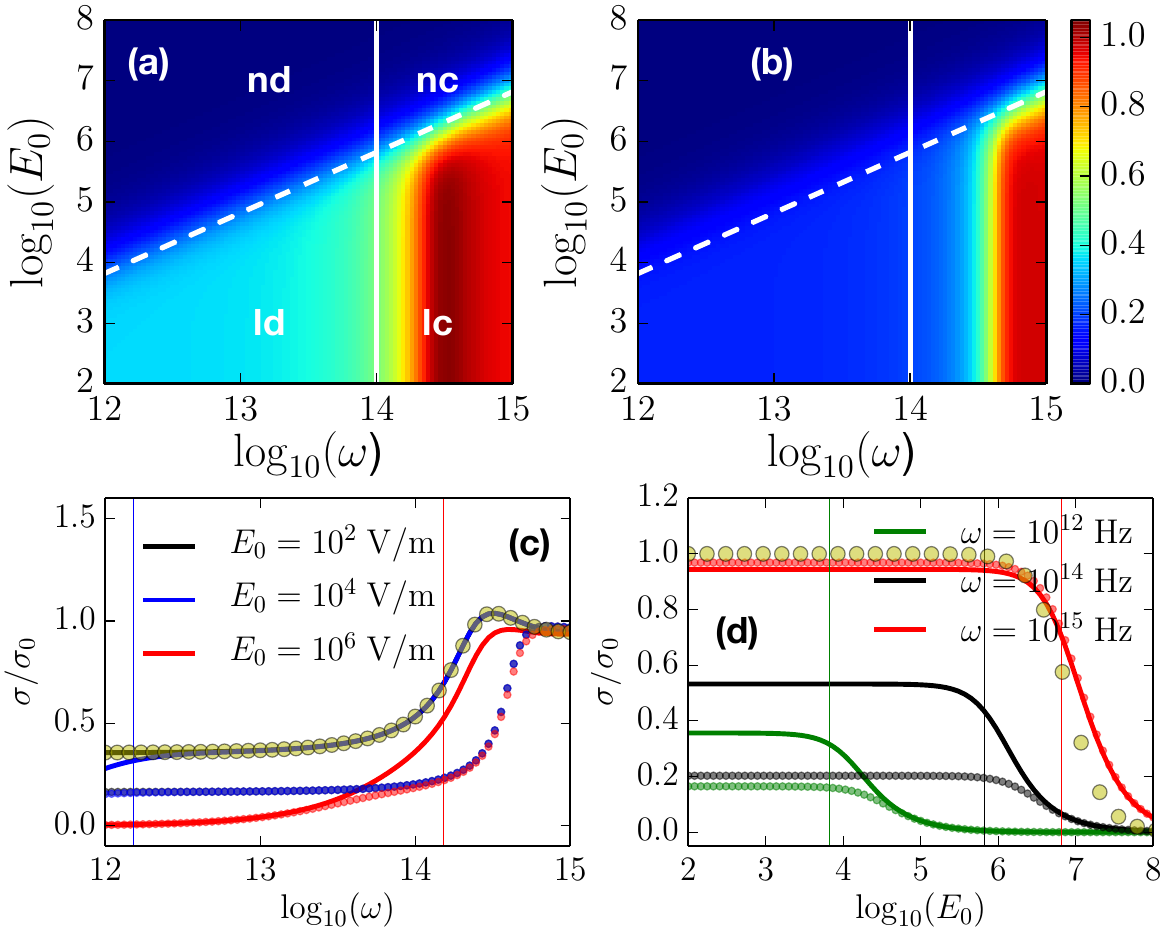}
\caption{Color plot of the optical conductivity as a function of frequency and the electric field strength of the incident laser beam for (a) gapped graphene ($\Delta =0.065{\rm eV} = \hbar \times 10^{14}$ Hz) with $\mu =0$  and (b) gapped and doped graphene with  $\mu = 2 \Delta $. As in the case of graphene,  the vertical solid white line at $\omega = \gamma_2$, and the dashed white line for $\zeta \equiv e v_{\rm F} E_I/(\hbar \omega \sqrt{\gamma_1 \gamma_2}) = 1$] divide the parameter space into 4 regimes: linear clean (marked 'lc'), non-linear clean ('nc'), non-linear dirty ('nd') and linear dirty ('ld'). Panel (c) show horizontal cuts from the upper two panels, i.e., the conductivity as a function of $\omega$ for different electric field strengths for both, the $\mu =0$ case (solid lines), and the doped case of $\mu = 0.13$ eV (dotted lines of the same color). The yellow circles show the excellent match of Eq.~\eqref{eq:ldmgr}, in the linear dirty limit, with the exact numerical results.
Panel (d) displays vertical cuts from the upper two panels, {\it i.e.}, the conductivity as a function of $E_{I}$ for different frequencies with solid lines for the $\mu =0$, and the dotted lines of the same color for $\mu = 0.13$ eV. Here the yellow circles show the excellent match of Eq.~\eqref{mgrcc}, in the non-linear clean limit, with the exact numerical results. In all the panels we have chosen $v_{\rm F} =10^6$ m/s, and $\{\gamma_1,\gamma_2\} = \{10^{12},10^{14}\}$ Hz.
\label{fig5}}
\end{figure}

In the {\it linear clean} limit ($\zeta \ll 1,~\gamma_2/\omega \ll1$), using Eq.~\eqref{eq:Md} in Eq.~\eqref{eq:aa}, we obtain the optical conductivity for massive graphene (with $g_s =g_v=2$) to be 
\be \label{nocmgk}
\sigma_{xx}^{\rm lc,m}(\omega) = \frac{g_sg_ve^2}{16\hbar}\left[1+ \frac{4\Delta^2}{\hbar^2\omega^2}\right]g(\omega,\max(|\mu|,\Delta),T),
\ee
where function $g(x)$ is defined in Eq.~\eqref{eq:g}. In the limiting case of $T \to 0$, $g(x) \to \Theta(x)$ and 
\be
\sigma_{xx}^{\rm lc,m}(\omega) = \frac{e^2}{4\hbar}\left[1+ \frac{4\Delta^2 }{\hbar^2\omega^2}\right]\Theta\left(\frac{\hbar\omega}{2} - \max(|\mu|,\Delta)\right).
\ee
The above expression clearly suggests that the chemical potential is not important if it lies inside the gap~($\mu < \Delta$). This is a direct consequence of unavailability of any phase space below the bandgap for optical excitations. Evidently in the $\Delta \to 0$ limit, we recover the corresponding optical conductivity expression for graphene. 

In the {\it nonlinear clean} limit ($\zeta \geq 1$ and $\gamma_2/\omega \ll 1$), we use Eq.~\eqref{eq:cc} to calculate the optical conductivity at finite temperature. For massive Dirac systems in two dimensions, we obtain, 
\be \label{mgrcc}
\sigma_{xx}^{\rm nc,m} = \frac{e^2 g(\omega,\alpha,T)}{2\hbar\zeta^2} \left[ 1 - \frac{1}{\sqrt{1+\zeta^2}}\left(1+ \frac{4\Delta^2\zeta^2}{\hbar^2\omega^2} \right)^{-1/2}\right].
\ee
In the limiting case of $\Delta \to 0$, Eq.~\eqref{mgrcc} reduces to Eq.~\eqref{eq:cd} as expected.

Next we consider the  {\it linear dirty} limit ($\zeta \ll 1$ and $\gamma_2/\omega \geq 1$) and with the help of Eq.~\eqref{eq:bb}, we arrive at the following expression in the $T \to 0$ limit,
\be\label{eq:ldmgr}
\sigma_{xx}^{\rm ld,m} = \frac{e^2\gamma_2}{4\pi\hbar\omega}\int_{\frac{2 \alpha}{\hbar}}^{\frac{2 \Lambda}{\hbar}}d\omega_{\bf k}\left[\frac{4 \hbar^{-2}\Delta^2 + \omega_{\bf k}^2}{\omega_{\bf k}
\left[(\omega_{\bf k}-\omega)^2 + \gamma_2^2\right]} - (\omega \to 0) \right],
\ee
where $\Lambda$ is the ultraviolet energy cutoff and $\alpha = {\rm max}\{|\mu|, \Delta\}$.  Evaluating Eq.~\eqref{eq:ldmgr} we obtain
\be \label{eq:ldmgr1}
\sigma_{xx}^{\rm ld,m}(\omega) = \frac{e^2}{4 \pi \hbar} \left[f_2\left(\omega,  2 \Lambda/\hbar \right) - f_2\left(\omega, 2 \alpha/\hbar \right) \right]~,  
\ee
where we have defined,
\bearr 
f_2(\omega, x) &\equiv& \left(1+y \right) \tan^{-1}\left(\frac{x-\omega}{\gamma_2}\right) - \frac{\gamma_2^2 - 4\hbar^{-2} \Delta^2}{2 \gamma_2 \omega} \ln[x^2 + \gamma_2^2 ]  \nonumber \\ 
& + &  \frac{\gamma_2(1-y)}{2\omega}  \log[(x-\omega)^2 + \gamma_2^2] - \frac{\omega y}{\gamma_2} \log(x) ~, \nonumber \\
\eearr
and $y \equiv 4 \hbar^{-2}\Delta^2/(\omega^2 + \gamma_2^2)$.
The finite temperature generalization of Eq.~\eqref{eq:ldmgr1} has to be calculated numerically. As a simple check we note that as $\Delta \to 0$, 
Eq.~\eqref{eq:ldmgr1} reduces to Eq.~\eqref{eq:ldgr1}. 

In the most general case, corresponding to the {\it nonlinear dirty} limit for massive graphene,  the dissipative part of the current density which arises only from the inter-band contribution, is given by 
${\bf \tilde J}_{\bf k}^{\rm m} = \left({A}_{{\bf k}}, {B}_{\bf k}\right)n^{\rm eq}_{\bf k}$, where 
\begin{widetext}
\bearr \label{eq:DeltaJ}
A_{\bf k} &=& \frac{e^2v_{\rm F}^2 E_0\gamma_{1}\left[ \Delta^2\gamma_{2}\cos\theta_0+ \hbar^2 v_{\rm F}^2 k^2 \gamma_{2}\sin(\phi_{\bf k}-\theta_0)\sin\phi_{\bf k} - \tau \Delta g_{\bf k}(\omega_{\bf k}-\omega)\sin\theta_0\right]}{\hbar \omega g_{\bf k}^2\left\lbrace \gamma_{1}\left[(\omega_{\bf
k}-\omega)^2+\gamma_{2}^2\right]+\gamma_{2}|\Omega_{\bf
k}^{cv}|^2\right\rbrace},\\ \label{eq:DeltaJ2}
B_{\bf k} &=& \frac{e^2v_{\rm F}^2 E_0\gamma_{1}\left[\tau \Delta g_{\bf k}(\omega_{\bf k}-\omega)\cos\theta_0- \hbar^2 v_{\rm F}^2 k^2\gamma_{2}\sin(\phi_{\bf k}-\theta_0)\cos\phi_{\bf k}+\Delta^2\gamma_{2}\sin\theta_0\right]}{\hbar \omega
g_{\bf k}^2\left\lbrace \gamma_{1} \left[(\omega_{\bf k}-\omega)^2+\gamma_{2\bf 
k}^2\right]+\gamma_{2}|\Omega_{\bf k}^{cv}|^2\right\rbrace} ~.
\eearr
\end{widetext}
However for this case,  the $\bf k$ integration of Eq.\eqref{eq:DeltaJ} has to be done numerically.

\begin{figure}[t!]
\centering
\includegraphics[width=0.99\linewidth]{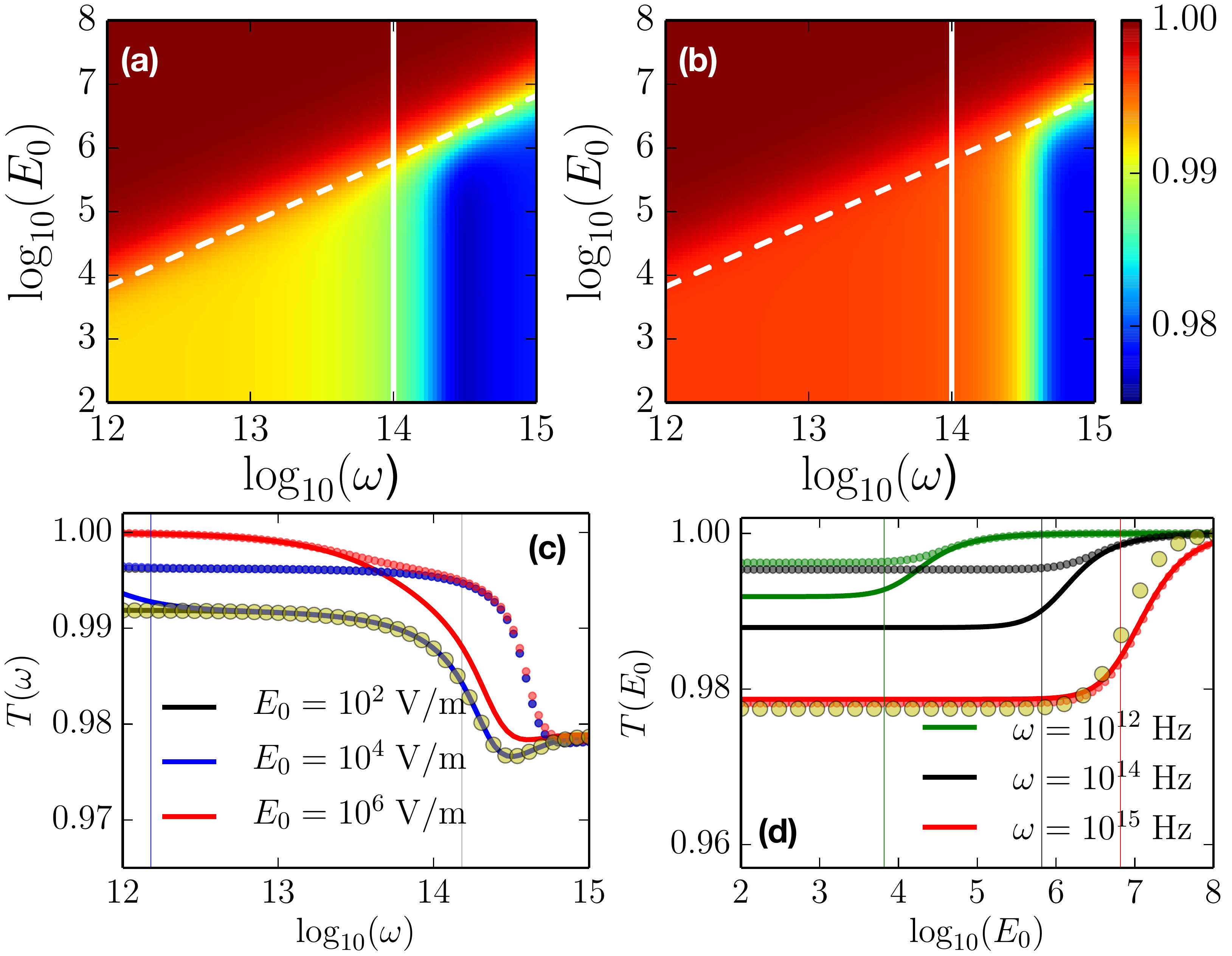}
\caption{Color plot of the nonlinear transmission as a function of frequency and the electric field strength of the incident laser beam for (a) pristine graphene with $\mu =0$  and (b) doped graphene with  $\mu = 0.13 $ eV (or equivalently $\mu/\hbar = 2 \times 10^{14}$ Hz).  Panels (c) and (d) display horizontal and vertical cuts, respectively with the solid lines corresponding to the $\mu = 0$ case, and the dotted lines representing the $\mu = 0.13$ eV. Other parameters are identical to that of Fig.~\ref{fig5}.
\label{fig4b}}
\end{figure}

The optical conductivity of massive graphene, in all four regimes is shown in Fig.~\ref{fig5}. As in the case of graphene, the $\zeta = 1$ line marks the boundary of the non-linear response regime, saturable absorption effects 
dominating beyond $\zeta > 1$. Note that the Kubo formula based result for the 
optical conductivity of massive graphene [Eq.~\eqref{nocmgk}], is valid only in the $\omega > \gamma_2$ and $\zeta \ll 1$ regime. The impact of non-linear optical conductivity on the transmission spectrum is highlighted in Fig.~\ref{fig4b}. 

Finally we note that in graphene $\sigma_{yx}(\omega) $ was zero for a $\hat x$ polarized light for each valley. This is not the case for massive graphene. We find that based on Eq.~\eqref{eq:DeltaJ2}, for a $\hat x$ polarized light, each valley has a finite $\sigma_{yx}(\omega)$, which in the clean linear response regime is given by 
\be \label{sigmaxy}
\sigma_{yx}(\omega) = \frac{2 \sigma_0}{\pi} \sum_\tau \frac{\tau \Delta_{\tau}}{\hbar \omega} \log \left(\frac{2 ~{\rm max}\{\Delta_\tau, \mu\}}{2 ~{\rm max}\{\Delta_\tau, \mu\} - \omega} \right)~, 
\ee 
where $\tau = + 1$ ($-1$) corresponds to the $K$ ($K'$) valley and $\Delta_\tau$ is the corresponding gap.  
Note that in Eq.~\eqref{sigmaxy} the sign of $\sigma_{yx}(\omega)$ in the $K$ valley turns out to be opposite to that of the $K'$ valley and if they have the same gap, the total $\sigma_{yx}(\omega)$ vanishes. 
However, if a valley asymmetry can be induced in graphene or other Dirac material (by breaking time reversal symmetry) [\onlinecite{Hunt1427}], leading to different band gap at the $K$ and $K'$ valleys ($\Delta_{K} \neq \Delta_{K'}$), then we can have a finite $\sigma_{yx}(\omega)$.

\section{Conclusion}
\label{sec7}
In this paper, we have present a unified formulation to calculate the non-linear optical conductivity for a generic two band system.   
Our model is based on a steady state solution of the optical-Bloch equations which yields an analytic expression for the population inversion and the inter-band coherence. A natural outcome of our model is the appearance of the dimensionless parameter $\zeta \propto E_0/\omega$, 
which quantifies the degree of optical non-linearity in the system, which was first pointed out by Mishchenko in the context of graphene [\onlinecite{Mishchenko}]. This implies that nonlinear saturation effects are stronger at lower frequencies for the same strength of the optical field strength. Furthermore, based on the parameter $\zeta$ and the coherence decay rate $\gamma_2$, any optical two band system can be said to be in one of the four regimes: (1) linear clean where $\zeta \ll1$, and $\gamma_2 \ll \omega$, (2) linear dirty  where $\zeta \ll1$, and $\gamma_2 \geq \omega$, (3) non-linear clean where $\zeta \geq 1$, and $\gamma_2 \ll \omega$, and (4) non-linear dirty where $\zeta \geq 1$, and $\gamma_2 \geq \omega$. These regimes present distinct signatures in the optical conductivity and the optical transmission and reflection spectrum.

Having established a general formulation for any two band system, we explicitly study the non-linear optical conductivity of graphene and massive graphene using the effective low energy Hamiltonian, and find analytic expressions for the optical conductivity in various regimes, reproducing the results for the clean case in the linear response regime.  We emphasize that the usually reported Kubo formula based results for the optical conductivity are generally valid only in the high frequency ($\omega  \gg \gamma_2$) and linear response ($\zeta \ll 1$) regimes. 

An obvious extension of this work is to include the electron-phonon and electron-electron interactions explicitly along  with the optical Bloch equation. This will provide a natural microscopic model for population inversion and decoherence decay rates, $\gamma_1$ and $\gamma_2$, which we have assumed to be constant in this paper. Along with this the effect of band bending, trigonal warping etc. can be included by considering a tight-binding model for the Hamiltonian as opposed to an effective low energy Hamiltonian, and that will also increase the validity of this formulation for a wide range of optical frequencies.  

\bibliography{refs_NLC}
\end{document}